\documentclass[global,twocolumn]{svjour}
\usepackage{graphics,amsmath}
\journalname{Applied Physics B}
\begin{document}
\title{A portable laser system for high precision atom interferometry experiments}
\author{Malte Schmidt\inst{1} \and Marco Prevedelli\inst{2} \and Antonio Giorgini\inst{3} \and Guglielmo M. Tino\inst{3} \and Achim Peters\inst{1}% etc
}                     % Do not remove
\institute{Humboldt-Universit\"at zu Berlin, Institut f\"ur
Physik, Newtonstra{\ss}e 15, 12489 Berlin, Germany,\\email:
malte.schmidt@physik.hu-berlin.de, Tel. +49-30-2093-4941, Fax.
+49-30-2093-4718 \and Dipartimento di Fisica, Universit\`a di
Bologna, Via Irnerio 46, 40127 Bologna, Italy \and Dipartimento di
Fisica e Astronomia and LENS, Universit\`a di Firenze-INFN, via
Sansone 1 Polo Scientifico, 50019 Sesto Fiorentino (Firenze),
Italy}
\date{Received: date / Revised version: date}
\maketitle
\begin{abstract}
We present a modular rack-mounted laser system for the cooling and
manipulation of neutral rubidium atoms which has been developed
for a portable gravimeter based on atom interferometry that will
be capable of performing high precision gravity measurements
directly at sites of geophysical interest. This laser system is
constructed in a compact and mobile design so that it can be
transported to different locations, yet it still offers
improvements over many conventional labora\-tory-based laser
systems. Our system is contained in a standard 19" rack and emits
light at five different frequencies simultaneously on up to 12
fibre ports at a total output power of 800 mW. These frequencies
can be changed and switched between ports in less than a
microsecond. The setup includes two phase-locked diode lasers with
a phase noise spectral density of less than 1 $\mu$rad/Hz$^{1/2}$
in the frequency range in which our gravimeter is most sensitive
to noise. We characterize this laser system and evaluate the
performance limits it imposes on an
interferometer.\\\end{abstract}

\section{Introduction}
\label{intro} Since first experimental demonstrations in 1991
\cite{Carnal1991,Riehle1991,Keith1991,Kasevich1992}, atom
interferometry has developed into a powerful tool for the ultra
precise measurement of accelerations and rotations. It is now used
in various laboratories for experiments in the fields of
fundamental physics \cite{Mueller2010,Dimopoulos2008} and
metrology \cite{Peters2001,Gustavson2000,Snadden1998}. In
principle, this new technique is also ideally suited for
high-accuracy field research such as gravity mapping, geophysics,
seismology or navigation \cite{Angelis2009} and could
substantially exceed that of classical gravi\-meters
\cite{Le2008}. However, due to the complexity of these experiments
they were so far confined to laboratory environments. Only in
recent years efforts have been undertaken to develop mobile atom
interferometers \cite{Stern2009,Cheinet2006} that might in future
versions also be used on satellite missions
\cite{Tino2007,Turyshev2007}. Our transportable high-precision
gravimeter GAIN (Gravimetric Atom Interferometer) is designed
specifically for geophysical on-site measure\-ments.

The working principle of a gravimetric atom interferometer has
been described in detail elsewhere \cite{Peters2001,Dubetsky2006}.
In short: An ensemble of laser-cooled neutral atoms (in our case
Rubidium 87) is prepared in a 3D Magneto-Optical Trap (MOT),
further cooled in optical molasses, and launched upwards. During
their parabolic flight, the atoms are subjected to three pulses
from counterpropagating laser beams, thereby inducing two-photon
Raman transitions that transfer them between the two hyperfine
ground states via a stimulated Raman process. The sequence
consists of one $\frac{\pi}{2}$-, one $\pi$-, and finally another
$\frac{\pi}{2}$-pulse, which constitute an atom optic beam
splitter, mirror and recombiner, respectively. Thereby the atomic
wave packet is split into two parts that travel on different
trajectories due to momentum transfer from the photons. At the
output of the atom interferometer the transition probability $P$
from one hyperfine state to the other is given by
$P=\frac{1}{2}(1+C\cos\Delta\Phi)$, where $\Delta\Phi$ is the
accumulated phase difference between wave packets and $C$ the
contrast of the measurement. This includes an acceleration
contribution of $\Delta\Phi_{g}=k_{\text{eff}}\,gT^2$ with $T$
being the time between two consecutive Raman pulses,
$k_{\text{eff}}=k_1+k_2$ the sum of the two individual
couterpropagating Raman laser wavenumbers $k_i$, and $g$ the local
gravitational acceleration.

\begin{figure*}
\centering\resizebox{0.9\textwidth}{!}{%
  \includegraphics{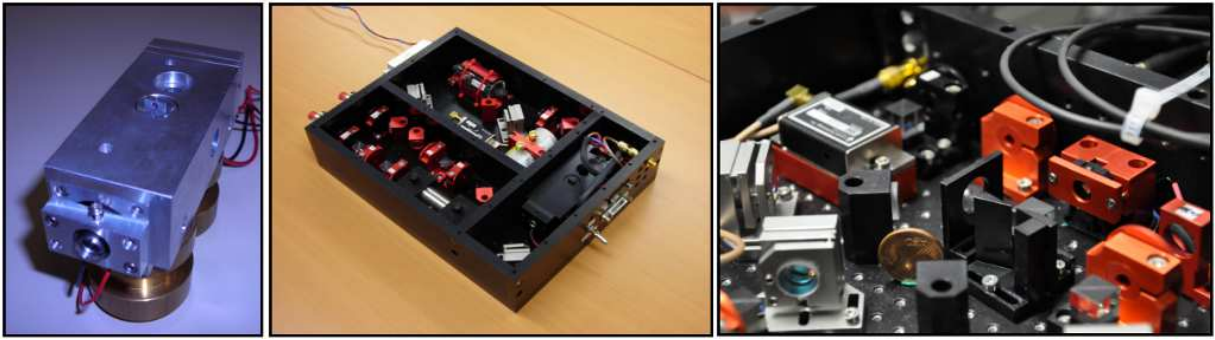}}\caption{Custom-made miniaturized optical mounts. Beam height for all
optics is 20 mm, round optical components have a diameter of
$0.5"$. Left to right: External Cavity Diode Laser, Reference
laser module, Raman laser module detail with a 5-Eurocent coin
($d=21$ mm) as reference} \label{optics}
\end{figure*}

In this paper, we describe in detail the realization of the atomic
gravimeter's laser system and characterize its subcomponents. The
sensor's physics package (i.e. main chamber and attached optics
and electronics, vibration isolation system, control system),
first high-precision gravity measurements as well as the
instrument's performance in the field will be discussed in future
publications.

\section{Concept}
\label{sec:concept}

Naturally, the laser system for a mobile atom interferometer has
to be mobile and compact as well. Our system is designed to be
``truckable'' (i.e. easily transportable by a small truck from one
gravity measurement site to the next), requiring a
rethermalization and readjustment time of less than a day. It will
operate at gravimetry reference points where the absolute gravity
value is measured in regular intervals \cite{Timmen2008}. These
points are usually selected to be inside of buildings with an
environment that is relatively stable, as conventional gravimeters
are sensitive to changes in their operating environment (although
harsher conditions can be found at more unusual measurement
sites). We can thus expect to typically encounter environments
with temperature variations of one or two Kelvin, low vibrational
noise level and no direct sun light. This is, however, still
considerably worse than conditions usually found in laboratory
cold atom experiments. Standard laser systems for these
experiments are not only quite complex, but also can rarely endure
significant mechanical vibrations, thermal fluctuations of even a
few Kelvin or electromagnetical noise, without losing laser
frequency locks or a significant decrease in optical power output.

A first step in solving these problems was the redesign from
scratch of almost all optical mounts, as standard laboratory
equipment rarely offers sufficient mechanical stability and is in
most cases simply too large for our purposes (Figure
\ref{optics}). Many of these mount designs are adapted from the
QUANTUS drop tower project \cite{Zoest2010} and have proven their
stability even under extreme accelerations of up to 50 $g$. We
have mounted all optics in four closed, compact modules with 1 cm
thick walls and a beam height of 20 mm for high ruggedness. Light
is transferred between the modules by means of polarisation
maintaining optical fibres. Except for the reference laser, which
is even smaller, each module's base area is 42 by 42 cm with a
height of a few centimeters. The base plate of two of these
modules is a custom made Thorlabs aluminium honeycomb breadboard
with a 1 cm grid of M3 threads which provides stability as well as
flexibility for possible future modifications. The other two
modules use 25 mm thick aluminum slabs as base plates with the
four walls and two additional divider walls mounted in a force-fit
such that under mechanical stress the assembled modules behave as
if manufactured from one piece. Fibre and electrical connections
are mounted on the walls of each module, the total height of each
module varies between 75 and 105 mm.

\begin{figure}
\centering\resizebox{0.45\textwidth}{!}{%
  \includegraphics{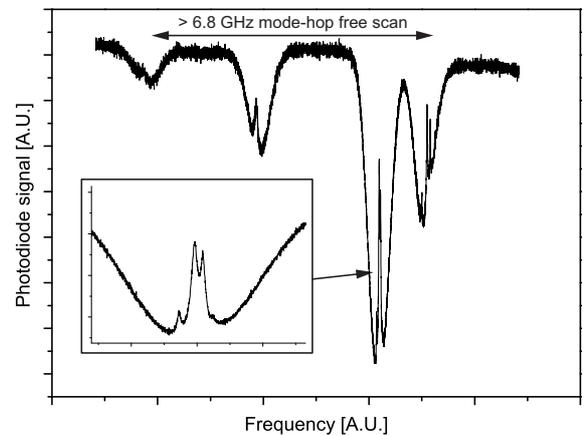}
} \caption{Mode-hop free scan over the complete $^{87}$Rb D2 line spectrum
using one of our ECDLs. Insert: $^{85}$Rb F=3 $\rightarrow$ F'
lines that we use to stabilise the laser.} \label{ecdlspectroscopy}
\end{figure}

As laser sources we have built compact external cavity diode
lasers (ECDL) that have been adapted from a design developed at
SYRTE \cite{Baillard2006} using Sharp GH0781\-JA2C laser diodes.
They include an interference filter as a wavelength selector that
is one order of magnitude less sensitive to angular misadjustments
than gratings found in conventional ECDL configurations which
makes it suitable for our application. The output power is up to
50 mW and the intrinsic linewidth less than 10 kHz. The cavity
length is 80 mm which gives a free spectral range (FSR) of
approximately 1.9 GHz. Control of the laser frequency is achieved
by tuning of either the laser diode current or the cavity length
(via a piezo). Applied separately, the mode-hop-free tuning range
is limited to 200 MHz or the FSR of 1.9 GHz, respectively.
However, by controlling both parameters simultaneously at a fixed
gain ratio, a mode-hop free tuning range of about 9 GHz and
thereby over more than the complete $^{87}$Rb D2 line was achieved
(Figure \ref{ecdlspectroscopy}).

\begin{figure}
\centering\resizebox{0.45\textwidth}{!}{%
  \includegraphics{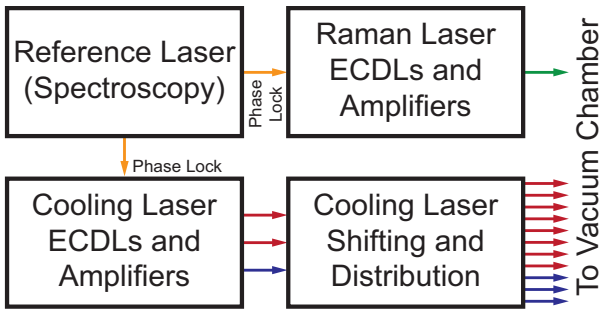}
} \caption{Modular concept of our laser system} \label{schematic}
\end{figure}

To provide the various optical frequencies required to operate an
atomic fountain interferometer and yet still maintain sufficient
flexibility for future enhancements like quasi-continuous
operation (many clouds of atoms in flight at the same time), a
total of five ECDLs is employed. They are organized in four
distinct modules: one reference laser module, two cooling laser
modules and one Raman laser module (Figure \ref{schematic}).

\begin{figure}
\centering\resizebox{0.35\textwidth}{!}{%
  \includegraphics{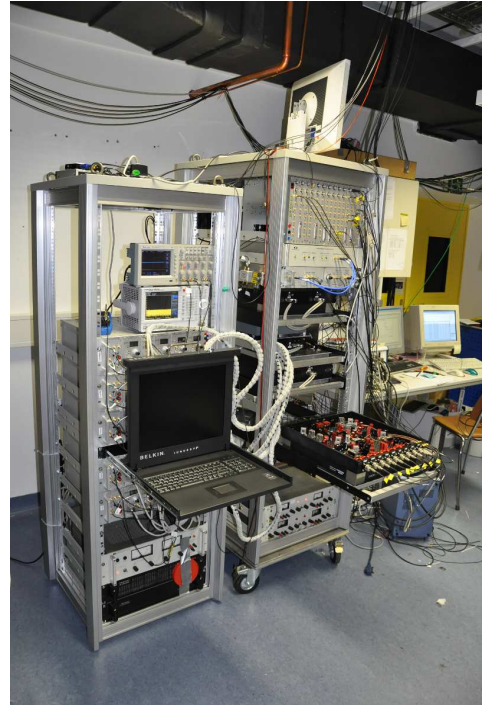}
} \caption{Photograph of laser system, cooling module 2 extended
on telescopic rails. In addition to the complete laser system,
these two racks also contain the gravimeter's computer, power
supplies, emergency backup batteries, control electronics and
diagnostic equipment} \label{photo1}
\end{figure}

The modules are mounted in a standard 19" electronics rack that
has been fitted with inflatable air bags between its main body and
its base plate which serve as a passive vibration isolation and
also as a shock absorber for transport over rough terrain. A test
transport over a snow-covered cobblestone surface has resulted in
a decrease of fibre coupling efficiency of less than 30 percent.
The rack is dimensioned such that it can fit through standard
doors and is thereby easily transportable to different locations
(Figure \ref{photo1}). For easy access, the laser modules are
mounted on telescopic rails. The complete laser system and its
control electronics could theoretically be mounted in just one
rack -- we chose, however, to also include other gravimeter
electronics such as computer control, a backup power system, power
supplies and diagnostic equipment which made a second rack
necessary and enabled us to move mechanically noisy equipment
(i.e. anything that includes cooling fans) away from the optics.

The dimensions of the complete system are $177\times60\times60$ cm$^3$
(computer and control electronics rack) plus $194\times80\times60$ cm$^3$
(laser rack), adding up to a total volume of $1.6$ m$^3$. Power
consumption is less than 1 kW.

\section{Reference Laser}
\label{sec:reference}

\begin{figure}
\centering\resizebox{0.35\textwidth}{!}{%
  \includegraphics{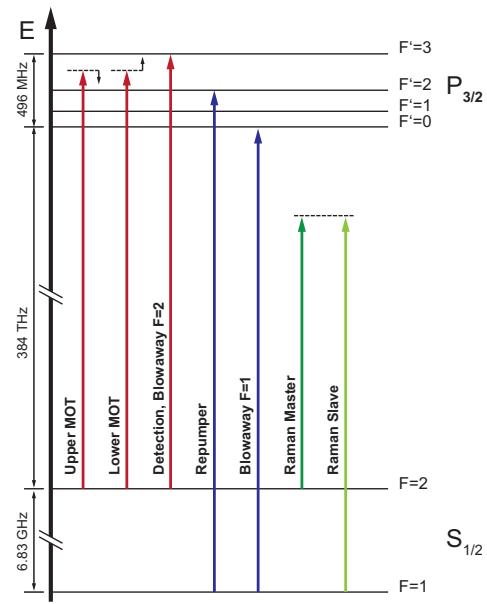}
} \caption{$^{87}$Rb D2 level scheme and laser frequencies required in
our setup} \label{levels}
\end{figure}

The reference laser module houses an ECDL that is stabilized 40
MHz below the $^{85}$Rb F=3 $\rightarrow$ F'=4 transition by
modulation transfer spectroscopy \cite{Supplee1994}. The control
signal is used to stabilize the laser wavelength via the the laser
diode's current controller. However, since this parameter is
limited to a mode-hop free tuning range of 200 MHz, a second
control path is employed: low-frequency components (below 100 Hz)
of the control signal are applied to both the laser diode current
as well as the piezo voltage at the fixed gain ratio determined to
give us a mode-hop free tuning range of 9 GHz (see above). This
way, slow drifts due to temperature variations or mechanical
stress are compensated and the laser stays locked over days at its
required frequency. In a laboratory test, the reference laser
stayed locked even when slowly changing the module's temperature
by 10 Kelvin over two hours. The reference laser is also
impervious to moderately strong hits of a metallic wrench to its
base plate. Comparing the system's open-loop and closed-loop
frequency responses using a network analyzer, we measured a
locking bandwidth of approximately 300 kHz (at 3dB below unity
gain). This bandwidth is limited by electronics and cable lengths
and not by the optical spectroscopy itself, as we reached a regime
in which the atoms are pumped quickly into the desired internal
states by using a pumping beam power of half a milliwatt with a
beam diameter of one millimeter.

\section{Cooling Laser System}
\label{sec:cooling}

An atomic fountain setup requires a variety of optical frequencies
in order to trap, cool, launch, select and detect the atoms, as
can be seen in Figure \ref{levels}. In our vacuum chamber, we
capture the atoms in a Magneto-Optical Trap (MOT). Since the atom
interferometer's sensitivity scales with the square root of both
the repetition rate and the number of atoms, we aim to trap as
many atoms as possible as quickly as possible. To achieve this, we
use a large MOT volume (beam diameter 35 mm) at high laser power
($>300$ mW). In order to launch the atoms, both upper and lower
MOT beams have to be detuned with respect to each other to achieve
a moving molasses configuration \cite{Kasevich1991}. Additionally,
a repumper is required, as well as blow-away and detection light
in both F=1$\rightarrow$F' and F=2$\rightarrow$F' frequency
classes. A total of 11 separate fibre output ports at different
laser powers and frequencies is required for full functionality.

Due to the large hyperfine ground splitting of 6835 MHz, light
from the F=1$\rightarrow$F' frequency class cannot easily (or with
high efficiency) be shifted to F=2$\rightarrow$F' by means of
acousto-optical modulators. Hence, we employ two ECDLs mounted in
the first cooling laser module, one for each frequency class. For
frequency stabilization, light from each ECDL is overlapped with
light from our reference laser on a fast photodiode (Hamamatsu
G4176-03) resulting in beat frequencies of about 5080 MHz and 1000
MHz, respectively. The signals frequencies are subsequently
divided down to about 19.8 MHz (factor 256) and 100 MHz (factor
10), respectively. Comparing the resulting signal's zero-crossings
to those of a stable Direct Digital Synthesizer (DDS) reference
frequency in a Hittite HMC440QS16G digital phase-frequency
detector (PFD) gives an error signal that we use to phase-lock the
beat signal onto the DDS reference at a bandwidth of 200 kHz. This
setup enables us to reach any desired laser frequency simply by
changing the DDS frequency. The F=2$\rightarrow$F' laser light is
split into two halves of 12 mW each that are used to seed two
Eagleyard tapered amplifiers (TPA-0780-01000), thereby amplifying
laser power to two times one Watt. Both of these high-power beams
mode-cleaned by an optical fiber at a coupling efficiency of only
between 50 and 60 percent due to imperfections in the tapered
amplifier's output laser beam profile. A third fibre is employed
for the F=1$\rightarrow$F' light (17 mW in-fibre).

\begin{figure*}
\centering\resizebox{\textwidth}{!}{%
  \includegraphics{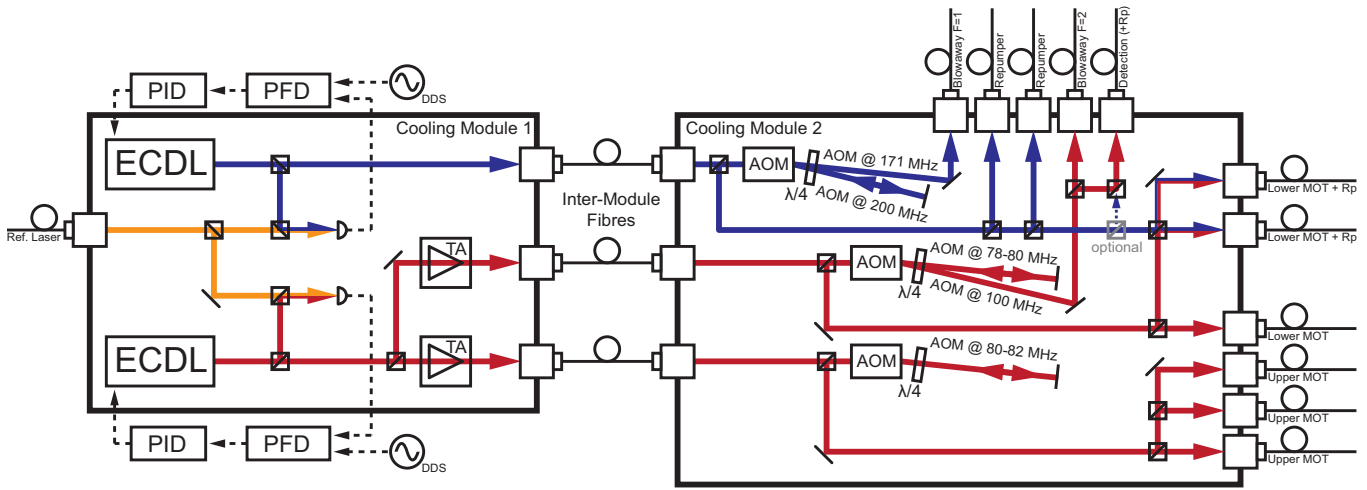}}\caption{Layout of Cooling laser modules} \label{coolsetup}
\end{figure*}

\begin{figure*}
\centering\resizebox{0.6\textwidth}{!}{%
  \includegraphics{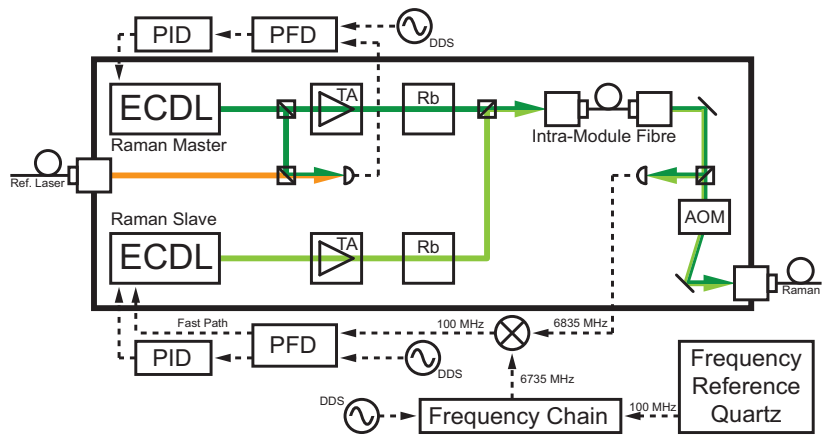}}\caption{Layout of Raman laser module} \label{ramansetup}
\end{figure*}

In the second cooling laser module the light is frequency shifted
and switched using acousto-optical modulators (AOMs), one for each
of these three beams. This enables separate frequency and
switching control of upper MOT, lower MOT and repumper beams. Not
all eleven output ports will have to be used simultaneously, as
for instance blow-away beams and MOT light are not required at the
same time. By changing the AOMs' frequencies, the first order
diffraction beam angle varies slightly. We make use of this effect
and hit different mirrors at different AOM frequencies, one of
which directs the light into one fibre, the other one reflects the
light back into the AOM for a double pass configuration.
Effectively, this enables us to switch between different fibre
outputs at different laser frequencies within less than a
microsecond without losing any light power at ports that are not
in use at any given moment. Since the most light power is needed
in the MOT phase of our experiment, the AOMs have been selected
and adjusted so that they work at their center frequency and
therefore peak efficiency of about 80 percent (single-pass) in
that configuration. For producing blow-away and detection beams,
less light power is sufficient, so we chose this configuration for
driving the AOMs far detuned from their center frequency which
resulted in single-pass efficiencies of 50 percent (driving an 80
MHz AOM at 100 MHz, Crystal Technology model 3080-125) and 60
percent (driving a 200 MHz AOM at 171 MHz, Crystal Technology
model 3200-121).

To avoid unwanted scattered light, additional mechanical shutters
(Sunex SHT934, switching time 1-2 ms) are used at each of the
eleven output ports. We have not observed any influence of the
shutters' mechanical noise on the fibre coupling efficiency. A
schematic of this setup is shown in Figure \ref{coolsetup}, the
total output power of this system in MOT configuration is six
times 60 mW cooling light plus 5 mW repumper light (in-fibre,
coupling efficiency larger than 80 percent).

\section{Raman Lasers}
\label{sec:raman}

In order to induce an optical Raman transition between the
hyperfine ground states of the atoms, a pair of two lasers with a
fixed phase relation and a frequency difference of 6.835 GHz (the
$^{87}$Rb ground state hyperfine splitting) is required to drive
the two-photon transition via an intermediate level, as seen in
Figure \ref{levels}. In this section, we will describe our Raman
laser system and its performance.

We employ two ECDLs (Figure \ref{ramansetup}) which emit light
that is amplified to 1 Watt using tapered amplifiers and then
overlapped. Rubidium vapour cells are employed to suppress
unwanted amplified spontaneous emissions on atomic resonances. An
AOM is used for fast switching and pulse-shaping of the Raman
pulses. Identical pulse-shaping on both laser beams is ensured by
an intra-module fibre common to both lasers where the light is
mode-cleaned before entering the AOM. The Raman master laser ECDL
is phase-locked to the reference laser using a setup similar to
the one employed in the cooling laser system, however the PFD we
use here is an Analog Devices ADF4108.

The Raman slave laser ECDL is stabilised in frequency and phase in
respect to the Raman master laser. Any phase noise between the two
Raman lasers will be imprinted onto the atoms and will therefore
directly limit the gravimeter's sensitivity. This is discussed in
detail in section \ref{sec:limit}. Accordingly, the requirements
in noise and locking bandwidth are much higher here than for the
locks of the Raman master or the two cooling lasers. For this
phase lock between the two ECDLs, light from both lasers is
overlapped on a fast photodiode that is placed behind the
intra-module fibre so that all noise sources that are not common
to both beams (i.e. anything before overlapping and mode-cleaning
done by the fibre) can be compensated for by the phase lock. The
resulting beat signal of 6835 MHz is mixed down using a stable
6735 MHz reference. The resulting 100 MHz signal is phase-locked
onto a DDS reference frequency using a Motorola MC100EP140 PFD. To
overcome bandwidth limitations and thereby residual phase noise
imposed by long cables and the laser diode current controller's
response time, an additional high frequency control path is
employed that modulates the ECDL laser diode's current directly
via a small signal N-channel FET as a voltage to current converter
and a lag-lead compensation network. Total cable lengths for this
fast path add up to less than a meter.

The 6735 MHz mixing-down frequency is generated by a frequency
chain that uses a Spectra Dynamics DLR-100 system as a frequency
reference. The DLR-100 includes an ultra-low noise 100 MHz quartz
that is locked to the 10th harmonic of a frequency-doubled 5 MHz
quartz for even lower phase noise at low frequencies. The 100 MHz
signal is multiplied to 6800 MHz and is then used to lock a
Dielectric Resonator Oscillator (DRO) to 6735 MHz. This frequency
chain is basically identical to the one described in
\cite{Le2008}.

\begin{figure}
\centering\resizebox{0.45\textwidth}{!}{%
  \includegraphics{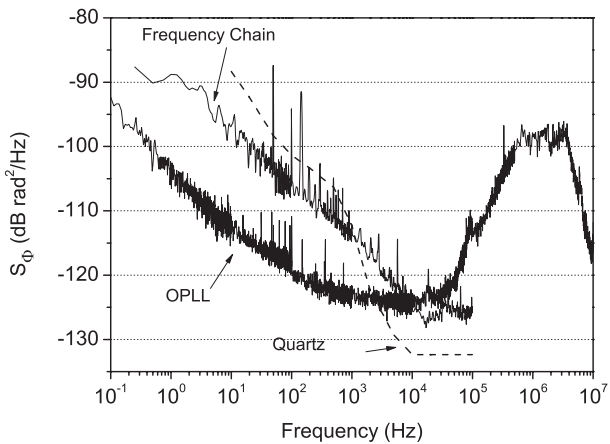}
} \caption{Phase noise spectral density for Raman laser OPLL,
frequency chain (data from \cite{Le2008}) and quartz (DLR-100) at
6.8 GHz} \label{psd}
\end{figure}

By comparing the downconverted signal with the 100 MHz DDS
reference using an independent mixer, we measured the phase noise
spectral density of this optical phase lock loop (OPLL), see
Figure \ref{psd}. Between 100 Hz and 60 kHz, where our
interferometer is most sensitive to noise (as shown in section
\ref{sec:limit}), it largely stays below a level of -120
dBrad$^2$/Hz (1 $\mu$rad/Hz$^{1/2}$). The achieved locking
bandwidth is slightly above 4 MHz. However, since in this
measurement any noise in either the frequency chain or the
reference quartz is cancelled out, both of these noise sources
have to be taken into account additionally in order to evaluate
the performance of the complete system: Thus, Figure \ref{psd}
also shows the phase noise spectral density of the frequency chain
as described and characterized in \cite{Le2008} and that of the
reference quartz system.

\begin{figure}
\centering\resizebox{0.45\textwidth}{!}{%
  \includegraphics{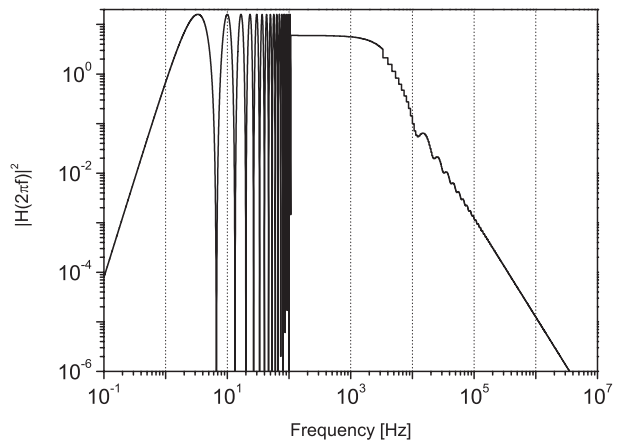}
} \caption{$|H(2\pi f)|^2$ calculated for $T$=150 ms and
$\tau$=100 $\mu$s, averaged after 16 oscillations}
\label{sens_function}
\end{figure}

\section{Gravimeter sensitivity}
\label{sec:limit}

In this section, we will calculate the limit to our gravimeter's
sensitivity due to phase noise from the Raman laser system. Of
course, these are not the only noise sources in our gravimeter, as
we are amongst others sensitive to mechanical vibrations and
detection noise. These noise sources will be evaluated in future
publications. In a setup like ours, the precision $\frac{\Delta
g}{g}$ with which we are able to measure local $g$ in a single
measurement is limited by the phase uncertainty $\Delta\Phi$ as
follows:

\begin{equation}
\frac{\Delta g}{g}=\frac{\Delta\Phi}{k_{\text{eff}}\,T^2g}
\end{equation}

The contribution of the Raman laser's power spectral density
$S_\Phi$ affects $\Delta\Phi$ via a transfer, or weighting
function, $|H(\omega)|^2$, i.e. $\Delta\Phi$ can be evaluated as

\begin{equation}
\Delta\Phi^2=\int^\infty_0|H(2\pi f)|^2S_\Phi(f)df
\end{equation}

In our setup we employ three Raman pulses: One $\frac{\pi}{2}$-pulse,
one $\pi$-pulse, and finally another $\frac{\pi}{2}$-pulse. Assuming
square Raman pulses of duration $\tau$, separated by time $T$, the
explicit form for $|H(\omega)|^2$ as derived in \cite{Cheinet2008}
is

\begin{equation}
\begin{array}{lcl}
|H(\omega)|^2&=&\left|-\frac{4\Omega\omega}{\omega^2-\Omega^2}\times\sin\left(\omega\frac{T+2\tau}{2}\right)\right.\\
&&\left.\times\left[\cos\left(\omega\frac{T+2\tau}{2}\right)+\frac{\Omega}{\omega}\times\sin\left(\omega\frac{T}{2}\right)\right]\right|^2\end{array}
\end{equation}

\begin{figure*}
\centering\resizebox{0.9\textwidth}{!}{%
  \includegraphics{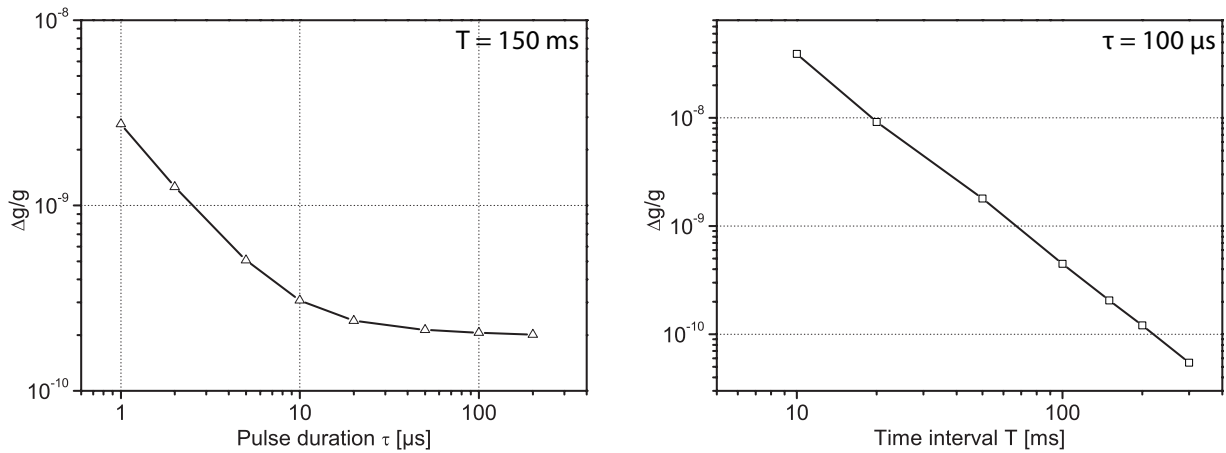}}\caption{Sensitivity limit given by Raman laser phase noise for
various values of $\tau$ and $T$} \label{tauvst}
\end{figure*}

with $\Omega=\pi/2\tau_R$ being the Rabi oscillation frequency of
the Raman transition. Due to a highly oscillatory behavior of
$H(\omega)$, however, after the 16th oscillation only the average
value is calculated (Figure \ref{sens_function}) in order to avoid
aliasing effects due to our limited data point spacing at higher
frequencies. Also of note is the band pass filter behavior of
$|H(2\pi f)|^2$ whose effective lower cutoff frequency scales with
$T^{-1}$, whereas the upper cutoff frequency scales with
$\tau^{-1}$.

To evaluate the limit that our laser system imposes on gravimeter
sensitivity, we calculate the root-mean-square of the spectra of
the three contributing sources of phase noise in our system (OPLL,
frequency chain, quartz) and multiply it with $|H(2\pi f)|^2$.
Integrating over the complete frequency spectrum, we obtain our
limits for $\Delta\Phi^2$ and consequently for $\frac{\Delta
g}{g}$. Assuming standard operating parameters for our gravimeter
of $T$=150 ms and $\tau$=100 $\mu$s, our single-shot sensitivity
will be limited to $\frac{\Delta g}{g}=1.93\times10^{-10}$ by
Raman laser phase noise.

\begin{figure}
\centering\resizebox{0.45\textwidth}{!}{%
  \includegraphics{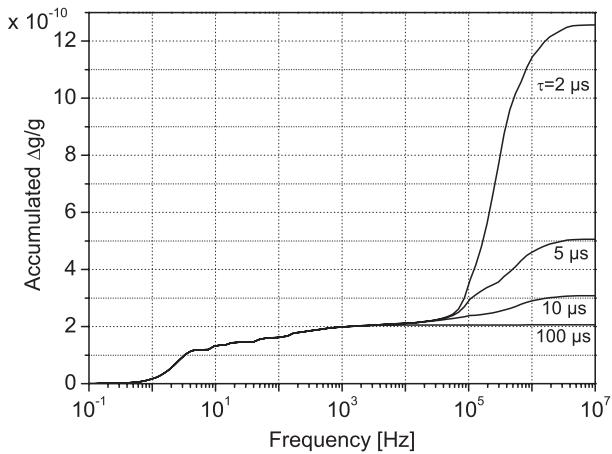}
} \caption{$\Delta g/g$ as a function of the upper limit in the
Raman laser power spectral density integration, displayed for
various pulse lengths $\tau$. $T=150$ ms is held constant.}
\label{acc_tau}
\end{figure}

For different $T$ and $\tau$, the band pass behavior of the
weighting function changes and so does our sensitivity to certain
phase noise frequencies. The resulting sensitivity limits are
shown in Figure \ref{tauvst}. To illustrate which frequencies are
the dominant contributors to these limit calculations, we have
plotted an accumulation integral for different pulse lengths
(Figure \ref{acc_tau}). In this diagram, $\frac{\Delta g}{g}$ is
plotted against $x$ in $\Delta\Phi^2=\int^x_0|H(2\pi
f)|^2S_\Phi(f)df$. As can be seen, our 4 MHz servo bump (see
Figure \ref{psd}) does not significantly contribute to the overall
gravimeter sensitivity at long pulse lengths $\tau$, at shorter
pulse lengths, however, the noise spectrum in the MHz range
becomes a dominating factor. In contrast, a variation of pulse
spacing $T$ instead of $\tau$ increases our sensitivity to
low-frequency phase noise which is dominated by our quartz
reference. This contribution is, however, not nearly as
significant as the fact that the sensitivity to $g$ scales with
$T^{-2}$, so we want to keep $T$ as long as possible in our
apparatus (see Figure \ref{tauvst}).

\section{Conclusion}
\label{sec:conclusion}

We have designed and built a laser system for atom interferometry applications that is mobile and robust, yet still offers improvements over many conventional laboratory-based systems. This system enables us to operate a highly precise atom interferometer outside of standard laboratory conditions and thereby opens up new possibilities for geophysical gravity measurements.

\section*{Acknowledgements}

This work is supported by the European Commission (FINAQS, Contr.
No. 012986-2 NEST), by ESA (SAI, Contr. No. 20578/07/NL/VJ) and by
ESF/DFG (Euro\-Quasar-IQS, DFG grant PE 904/2-1). We further thank
LNE-SYRTE for design and construction of the frequency chain and
the QUANTUS team for a fruitful collaboration.

\end{document}